\documentclass[aps,prd,showpacs,twocolumn,10pt,nofootinbib,floatfix,preprintnumbers]{revtex4}
\usepackage{amssymb}
\usepackage{amsmath}
\usepackage[dvips]{color}
\usepackage{graphicx}

\unitlength=1.2mm \preprint{JLAB-THY-07-761}
\begin{document}
\newcommand{\tr}{\mbox{tr}\,}
\newcommand{\Dslash}{{\mathchoice
    {\Dslsh \displaystyle}%
    {\Dslsh \textstyle}%
    {\Dslsh \scriptstyle}%
    {\Dslsh \scriptscriptstyle}}}
\newcommand{\Dslsh}[1]{\ooalign{\(\hfill#1/\hfill\)\crcr\(#1D\)}}
\newcommand{\leftvec}[1]{\vect \leftarrow #1 \,}
\newcommand{\rightvec}[1]{\vect \rightarrow #1 \:}
\renewcommand{\vec}[1]{\vect \rightarrow #1 \:}
\newcommand{\vect}[3]{{\mathchoice
    {\vecto \displaystyle \scriptstyle #1 #2 #3}%
    {\vecto \textstyle \scriptstyle #1 #2 #3}%
    {\vecto \scriptstyle \scriptscriptstyle #1 #2 #3}%
    {\vecto \scriptscriptstyle \scriptscriptstyle #1 #2 #3}}}
\newcommand{\vecto}[5]{\!\stackrel{{}_{{}_{#5#2#3}}}{#1#4}\!}
\newcommand{\vdot}{\!\cdot\!}

\bibliographystyle{apsrev}

\title{First Calculation of Hyperon Axial Couplings from Lattice QCD}

\author{Huey-Wen Lin}
\email{hwlin@jlab.org} \affiliation{Thomas Jefferson National
Accelerator Facility, Newport News, VA 23606
}
\author{Konstantinos Orginos}
\email{kostas@wm.edu} \affiliation{Thomas Jefferson National
Accelerator Facility, Newport News, VA 23606\\
Department of Physics, College of William and Mary, P.O.Box 8795,
Williamsburg, VA 23187
}

\date{Dec. 7, 2007}
\begin{abstract}
In this work, we report the first lattice calculation of  hyperon axial couplings, using the 2+1-flavor MILC configurations and domain-wall fermion valence quarks. Both the $\Sigma$ and $\Xi$ axial couplings are computed  for the first time in lattice QCD. In particular we find that $g_{\Sigma\Sigma} = 0.450(21)_{\rm stat}(27)_{\rm syst}$ and $g_{\Xi\Xi} = -0.277(15)_{\rm stat}(19)_{\rm syst}$.
\end{abstract}

\maketitle
\section{Introduction}
In recent years lattice QCD calculations have demonstrated remarkable progress in computing hadron properties from first principles. In particular, hadron structure has been a major focus of the lattice QCD community. Recently the nucleon axial coupling $g_A$ has been computed~\cite{QCDSFga,Edwards:2004sx} with good precision and has been shown to be in agreement with the very well known experimental result within the systematic and statistical errors of the calculation. The success of such calculations motivates us to study the axial couplings for all the octet baryons. The experimental knowledge of such couplings is not as good as in the case of $g_A$ and in certain cases, such as the axial couplings $g_{\Sigma\Sigma}$ and $g_{\Xi\Xi}$, their values  are not known experimentally, and theoretical estimates are rather imprecise.

The hyperon axial couplings are important parameters entering the low-energy effective field theory description of the octet baryons. At the leading order of SU(3) heavy baryon chiral perturbation theory, these coupling constants are linear combinations of the universal coupling constants $D$ and $F$, which enter  the chiral expansion  of  every baryonic quantity, including masses and scattering lengths. These coupling constants are needed in the effective field theory description of both the non-leptonic decays of hyperons, and the hyperon-nucleon and hyperon-hyperon scattering phase shifts~\cite{Beane:2003yx}. Hyperon-nucleon and hyperon-hyperon  interactions are essential  in understanding the physics of neutron stars where hyperon and kaon production may soften the equation of state of  dense hadronic matter.

Studying the octet baryon axial couplings on the lattice is no more complicated than computing the nucleon axial coupling $g_A$.
The lack of experimental information in cases such as $g_{\Sigma\Sigma}$ and $g_{\Xi\Xi}$ gives us the opportunity to make predictions using lattice QCD. Previously, there have been two attempts to determine these coupling constants using different theoretical approaches: chiral perturbation theory~\cite{Savage:1996zd} and the large-$N_c$ limit~\cite{Dai:1995zg,FloresMendieta:1998ii}. M.~J.~Savage et~al.~\cite{Savage:1996zd} use chiral perturbation theory to work out the one-loop corrections due to SU(3) symmetry breaking, and predict $0.35 \leq g_{\Sigma\Sigma} \leq 0.55$. Similarly, $0.18 \leq -g_{\Xi\Xi} \leq 0.36$, after taking the SU(3) limit, $g_{\Sigma\Sigma}=F$, which is 0.5 at the leading order. Using the large-$N_c$ approach, Dai et~al.~\cite{Dai:1995zg} and Flores-Mendieta et~al.~\cite{FloresMendieta:1998ii} predict a range of $0.30 \leq g_{\Sigma\Sigma} \leq 0.36$ and $0.26 \leq -g_{\Xi\Xi} \leq 0.30$. Both approaches give very loose bounds on the values of these coupling constants; hence, a lattice QCD calculation has the opportunity to make a substantial improvement.

Lattice QCD calculations can now provide much more stringent theoretical estimates of these axial couplings. The various systematic errors that enter such calculations can be controlled, giving us the ability to compute the predictions of QCD very precisely. The systematic errors that we need to control are due to the finite lattice spacing, the finite volume and chiral extrapolations. Lattice calculations are performed on a discrete space-time in a finite volume, using Monte Carlo integration to directly evaluate the path integral; continuum physics is recovered by taking the lattice spacing to zero ($a \rightarrow 0$) and the volume to infinity ($V \rightarrow \infty$). In addition, using current computer resources, we cannot yet calculate at the physical pion mass. Using chiral perturbation theory and calculations at multiple heavier pion masses which are more affordable, we can extrapolate quantities of interest to the physical limit. Such calculations also help to determine the low-energy constants of the chiral effective theory and allow us to study the quark-mass dependence of our observables.

The structure of this article is as follows: In Sec.~\ref{sec:parameters}, we describe our lattice setup, operators and the  parameters of the calculation. In Sec.~\ref{sec:hyperonAxial}, we discuss the chiral extrapolation, and finally, our conclusions are presented in Sec.~\ref{sec:conclusion}. Our preliminary results can be found in Ref.~\cite{Lin:2007gv}.


\section{The Lattice Calculation}\label{sec:parameters}
The axial coupling constants are defined as the zero momentum transfer limit of the axial form factor $G_A(q^2)$ that parametratizes the matrix element  $\langle B | {\cal O}| B\rangle$ where $B$ is one of the baryons $N$, $\Sigma$, $\Xi$, and ${\cal O}$ is the local axial current ($\overline{q}\gamma_\mu \gamma_5 q$). This  matrix element takes the form \begin{eqnarray}\label{eq:cont_ME}
\langle B\left|A_{\mu}(q)\right|B\rangle
&=& {\overline u}_{B}(p^\prime)\left[\gamma_{\mu}\gamma_5 G_A(q^2) +
\gamma_5 q_{\nu} \frac{G_P(q^2)}{2M_{B}}
\right]\nonumber \\
&& \times u_{B}(p) e^{-iq\cdot x},
\end{eqnarray}
where $u_B$ is the Dirac spinor, $G_P$ is the induced pseudoscalar
form factor and $q$ is the momentum transfer. The axial charge for baryon $B$ is defined as $g_{A,BB} = G_A (q^2=0)$.

On the lattice, we can extract the matrix element $ \langle B\left|A_{\mu}\right|B\rangle^{\rm lat}$ by calculating a three-point function using zero initial and final momentum for the baryon states.
This matrix element needs to be renormalized, because we use the local axial current in order to simplify our calculation. The renormalization constant $Z_A$ can be easily computed through two-point meson correlation functions. Details of our lattice formulation and methods can be found in Refs.~\cite{Edwards:2005ym,Sasaki:2003jh}.

For the rest of the paper, we will define  the renormalized zero mometum transfer matrix elements  as following: for the nucleon $g_A = Z_A \langle N\left|A_{\mu}\right|N\rangle^{\rm lat}$, the $\Sigma$ $g_{\Sigma\Sigma} = \langle \Sigma \left| A_{\mu}\right|\Sigma\rangle^{\rm lat}/2$ (the factor of 2 is coming from a Clebsch-Gordan coefficient so that $g_{\Sigma\Sigma} = F$ in the SU(3) limit)
and for the $\Xi$  $g_{\Xi\Xi} = Z_A \langle \Xi \left|A_{\mu}\right|\Xi\rangle^{\rm lat}$.

In this calculation, we use (improved) staggered fermion action (asqtad)~\cite{Kogut:1975ag,Orginos:1998ue,Orginos:1999cr} for the sea quarks and domain-wall fermions (DWF)~\cite{Kaplan:1992bt,Kaplan:1992sg,Shamir:1993zy,Furman:1994ky} for the valence sector.  This way we take advantage of the publicly available gauge configurations generated by MILC collaboration, only having to compute the valence quark propagators needed for the matrix elements. In addition, domain-wall fermions used in the valence sector are automatically $O(a)$ improved and have chiral and flavor symmetry which simplifies operator mixing, renormalization, and chiral extrapolation at finite lattice spacing~\cite{Bar:2005tu,Chen:2005ab,Aubin:2006hg,Chen:2007ug}, making them particularly well suited for the purpose of our calculation.
This mixed action approach has been successfully employed by LHPC and NPLQCD for computations of nucleon matrix elements~\cite{Renner:2004ck,Edwards:2005kw,Orginos:2006zz,Hagler:2007xi} as well as scattering lengths, decay constants~\cite{Beane:2005rj,Beane:2006gj,Beane:2006mx,Beane:2006kx,Beane:2006gf,Orginos:2007zz,Beane:2007uh,Beane:2007xs} and masses~\cite{Beane:2006fk,Beane:2006pt,Edwards:2006zz}~\footnote{Staggered fermions come in four ``tastes'' which must be removed by taking fractional powers of the fermionic determinant. Although no theoretical proof of the validity of this approach exists, there is significant evidence in the literature that supports the conjecture that the continuum limit of such formulation is QCD. For a recent review see Ref.~\cite{Sharpe:2006re} and references therein.}.

The gauge configurations are generated with 2+1 flavors of staggered fermions (configuration ensembles generated by the MILC collaboration~\cite{Bernard:2001av}). The pion mass ranges from  350 to 750~MeV in a lattice box of size 2.6~fm.
The gauge fields that enter the domain-wall fermion action are hypercubic smeared to improve the chiral symmetry, and gauge invariant Gaussian smearing  has been used for the interpolating operators to improve the signal (see~\cite{Renner:2004ck,Edwards:2005kw,Hagler:2007xi}  for details). The source-sink separation is fixed at 10 time units. The number of configurations used from each ensemble ranges from 300 to 600.
Our results  are presented in Table~\ref{tab:numbers}.

\begin{table}
\begin{center}
\begin{tabular}{c|ccccc}
\hline\hline
 	               & m010         &  m020        &  m030       &
    m040       & m050\\
\hline
$m_\pi$ (MeV) & 354.2(8)  & 493.6(6)  &  594.2(8) & 685.4(19) & 754.3(16)\\
$m_\pi/f_\pi$      & 2.316(7) 	  & 3.035(7)     & 3.478(8)    &
	3.822(23)  & 4.136(20)\\
$m_K/f_\pi$ 	   & 3.951(14) 	  & 3.969(10)    & 4.018(11)   &
	4.060(26)  & 4.107(21)\\
confs &  612 & 345 & 561 & 320 & 342 \\
$g_{A,N}$              &  1.22(8) & 1.21(5) &  1.195(17) & 1.150(17)&   1.167(11)\\
$g_{\Sigma\Sigma}$ & 0.418(23) &  0.450(15) &  0.451(7) & 0.444(8) &  0.453(5)\\
$g_{\Xi\Xi}$       & $-$0.262(13) & $-$0.270(10) & $-$0.269(7) &
 	$-$0.257(9)& $-$0.261(7)\\
\hline\hline
\end{tabular}
\end{center}
\caption{Results of our calculation}\label{tab:numbers}
\end{table}

\section{Hyperon Axial Coupling Constants}\label{sec:hyperonAxial}

\subsection{SU(3) symmetry breaking}\label{subsec:SU3}

\begin{figure}
\includegraphics[width=0.5\textwidth]{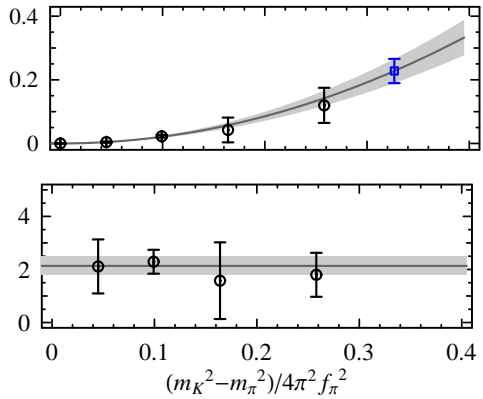}
\caption{(Top) The SU(3) symmetry breaking measure $\delta_{\rm SU(3)}$. The circles are the measured values at each pion mass, the square is the extrapolated value at the physical point, and the shaded region is the quadratic extrapolation and its error band. \\
(Bottom) $\delta_{\rm SU(3)}/x^2$ plot. Symbols as above, but the band is a constant fit.
}\label{fig:SU(3)-breaking}
\end{figure}

One way to probe SU(3) symmetry breaking in the axial couplings  is to monitor the quantity $\delta_{\rm SU(3)}$, defined as
\begin{eqnarray}
\delta_{\rm SU(3)} &=& g_{\rm A}- 2.0\times g_{\Sigma\Sigma} +g_{\Xi\Xi} = \sum_n c_n x^n;
\label{eq:SU3break}
\end{eqnarray}
where $x$ is ${(m_K^2-m_\pi^2)}/{(4\pi f_\pi^2)}$. Figure~\ref{fig:SU(3)-breaking} shows $\delta_{\rm SU(3)}$ as a function of $x$. Note that the value increases monotonically as we go to lighter pion masses. Our lattice data suggest that a $\delta_{\rm SU(3)} \sim x^2$ dependence is strongly preferred, as the plot of $\delta_{\rm SU(3)}/{x^2}$ versus $x$ in Fig.~\ref{fig:SU(3)-breaking} also demonstrates. A quadratic extrapolation to the physical point gives 0.227(38), telling us that SU(3) breaking is roughly 20\% at the physical point, where $x = 0.332 $ using the PDG values~\cite{PDBook} for $m_\pi^+$, $m_K^+$ and $f_{\pi^+}$. We compare the result of heavy baryon SU(3) chiral perturbation theory~\cite{Detmold:2005pt}  for  $\delta_{\rm SU(3)}$ as a function of $x$, and we find that the coefficient of the the linear term in Eq.~\ref{eq:SU3break} does not vanish. This implies that an accidental cancelation of the low-energy
constants is responsible for this behavior.

\subsection{Chiral extrapolation}\label{subsec:HBXPT}

Ideally, we should adopt a chiral extrapolation that correctly describes the discretization errors due to finite lattice spacing as we extrapolate the axial couplings from the pion masses we used in our calculation to the physical point. However, such a calculation does not exist in the literature. Here we consider  the next-best available option, which is continuum chiral perturbation theory.

W.~Detmold and C.-J.~D.~Lin worked out the forward twist-two matrix element extrapolation at one-loop order with finite-volume corrections~\cite{Detmold:2005pt}. From their work, we see that finite-volume corrections contribute at the order of $10^{-4}$ or less for the parameters of our calculation. Therefore, we will only implement the infinite-volume chiral perturbation theory and further simplify the formulation by taking $m_{\rm val}=m_{\rm sea}$. In these formulae, there are total eight parameters to be determined: three SU(3) coupling constants ($C$, $D$, $F$), and five other low energy constants ($\Delta c_n$, $\Delta \alpha_n$, $\Delta \beta_n$, $\Delta \gamma_n$, $\Delta \sigma_n$). We replace three mass splittings between the decuplet and octet with their lattice-measured values and replace $f$ and the chiral perturbation theory scale ($\mu$) with the pseudoscalar decay constants calculated from the same lattice and actions; a similar strategy was adopted by NPLQCD and LHPC~\cite{Beane:2005rj,Beane:2006gj,Beane:2006mx,Beane:2006kx,Beane:2006gf,Orginos:2007zz,Beane:2007uh,Beane:2007xs,Edwards:2006qx}. It has also been  shown  that such an approach simplifies the mixed action chiral perturbation theory formulas~\cite{Chen:2005ab,Chen:2007ug}. We perform a simultaneous fit among multiple axial coupling constants without making further assumptions. We attempted to do simultaneous fitting among all three $g_A$, $g_{\Sigma\Sigma}$ and $g_{\Xi\Xi}$, and found the $\chi^2/{\rm dof}$ for such a fit is of the order of $10^2$. If we limit ourselves to the lightest three pion masses, we can reduce $\chi^2/{\rm dof}$ to the order of $10$. Although under certain plausible assumptions for the values of some low energy constants, the individual axial couplings can be fitted with reasonable $\chi^2/{\rm dof}$, the combined fit seems to fail to describe the data. The conclusion we derive from this is that SU(3) heavy baryon chiral perturbation theory fails to describe the lattice data. The pion and kaon masses used in our calculation are probably outside the range of validity of this order of the perturbative expansion. It is possible that such a unified description of the axial couplings  requires the next order in chiral perturbation theory or smaller pion masses. However, since the physical strange quark mass is rather large, it seems that SU(3) heavy baryon chiral perturbation theory might not be useful in helping extrapolate the lattice data to the physical point. Here we should not forget that one of the reasons
for our failure to fit with continuum heavy baryon SU(3) chiral perturbation theory formulas could be  discretization errors due to taste symmetry breaking in the sea sector. However, such errors should be small; they are further suppressed to NLO in the expansion since the valence sector domain-wall fermions retain chiral symmetry. Hence we do not expect that finite lattice spacing corrections will change our conclusion.

\subsection{Simple chiral forms}\label{subsec:SU3-inspired-XPT}

\begin{figure}[t]
\includegraphics[width=0.5\textwidth]{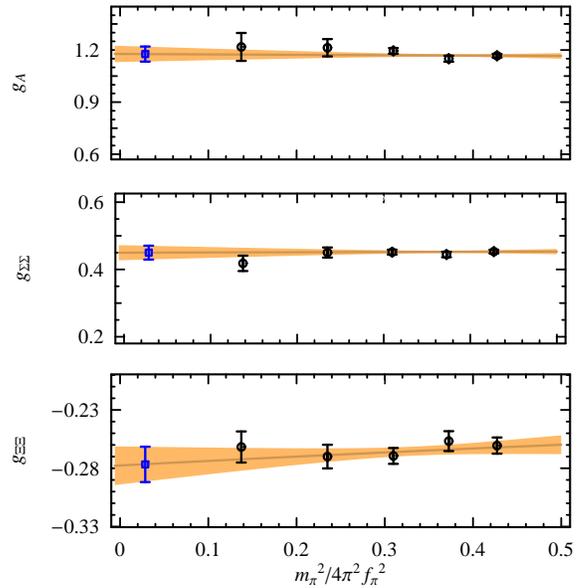}
\caption{Lattice data (circles) for $g_{A}$, $g_{\Sigma\Sigma}$ and $g_{\Xi\Xi}$ and chiral extrapolation (lines and bands). The square is the extrapolated value at the physical point.
}\label{fig:hyperonAxial}
\end{figure}

Since the continuum chiral extrapolation fails to describe our data, we take a step back and expand the axial couplings in terms of the  SU(3) breaking parameter $x=(m_K^2-m_\pi^2)/(4\pi^2f_\pi^2)$ as follows:
\begin{eqnarray}\label{eq:SU3-Form}
g_A &=&  D+F + \sum_n C_N^{(n)} x^n \nonumber\\
g_{\Sigma\Sigma} &=&  F + \sum_n C_\Sigma^{(n)} x^n\nonumber\\
g_{\Xi\Xi} &=&  F-D + \sum_n C_\Xi^{(n)} x^n\,.
\end{eqnarray}
This form reduces to the known SU(3)-symmetric limit where the axial couplings are simple linear combinations of $F$ and $D$. In addition, not all the $C_B^{(n)}$ (with $B \in \{N, \Sigma, \Xi\}$) are independent parameters. In Sec.~\ref{subsec:SU3} we find that the constraint  $C_N^{(1)}-2C_\Sigma^{(1)}+C_\Xi^{(1)}=0$ is preferred by the data. This suggests a 4-parameter fit to $n=1$ order or a 7-parameter fit to $n=2$ order if we ignore possible pion mass dependence of $F$ and $D$. In order to have a reasonable fit form  with the smallest number of parameters we will focus on the $n=1$ case. Figure~\ref{fig:hyperonAxial} shows our lattice data as a function of $(m_\pi/f_\pi)^2$ with the corresponding chiral extrapolation; the band shows the jackknife uncertainty.

The $\chi^2/{\rm dof}$ is 0.83 and the linear fit parameters are very poorly determined
as $C_N^{(1)}=0.02(13)$ and $C_\Sigma^{(1)}=-0.01(6)$. This is not surprising, since a small slope is seen in all three axial charges. At the physical pion point, we find the nucleon axial charge is $g_{A,N}=1.18(4)$; this is is consistent with what LHPC obtained, 1.23(8), from SU(2) chiral perturbation extrapolation~\cite{Edwards:2006qx}. The extrapolated coupling constants $g_{\Sigma\Sigma}$ and $g_{\Xi\Xi}$ are $0.450(21)$ and $-0.277(15)$ respectively. These numbers are consistent with the existing predictions from chiral perturbation theory~\cite{Savage:1996zd} and large-$N_c$ calculations but have much smaller errors. The low-energy chiral parameters are $D=0.715(6)$ and $F=0.453(5)$, which are not consistent with the recent determination of $D = 0.804(8)$ and $F = 0.463(8)$  using semileptonic decay data and assuming SU(3) symmetry~\cite{Cabibbo:2003cu}. However, the two calculations do agree within the range of the SU(3) breaking effect we observed in the previous section.

Since we know that the SU(3) breaking in Eq.~\ref{eq:SU3break} is quadratic in $x$, we expect that one needs to go at least to $n=2$ in order to capture this effect. Hence, when taking the $n=2$ expansion, it is not surprising to see the $\chi^2/{\rm dof}$ improves to 0.57. However, the coefficients $c_B^{(n)}$ remain poorly determined (most of them are consistent with zero within the errorbar). Note that $C_N^{(2)}-2C_\Sigma^{(2)}+C_\Xi^{(2)}=1.9(6)$ is consistent with what we found for the curvature of $\delta_{\rm SU(3)}$ in Sec.~\ref{subsec:SU3}. The final results, $D=0.711(7)$ and $F=0.452(5)$, are consistent with the $n=1$ case but are better determined. The axial couplings in this case are $g_A= 1.28(6)$, $g_{\Sigma\Sigma}= 0.39(6)$ and $g_{\Xi\Xi}= -0.24(4)$. The discrepancy between the results of $n=1$ and $n=2$, which is at one standard deviation, might be taken as an indication of the systematic error of such extrapolations. However, it is hard to make an honest determination of such systematic error without further study at lighter pion masses and higher statistics.

Finally, one can consider that $D$ and $F$  have $m_\pi^2$ dependence as
\begin{eqnarray}\label{eq:D-F}
&& D =  D_0 + c_D m_\pi^2/(4\pi^2f_\pi^2), \,\,\, F =  F_0 + c_F m_\pi^2/(4\pi^2f_\pi^2)\nonumber,
\end{eqnarray}
where $D_0$ and $F_0$  are the chiral limit axial couplings. Such fit forms have more parameters than we can possibly determine with our data. To better understand how strong this dependence is, let us assume $C_{B}^{(n)}=0$ in Eq.~\ref{eq:SU3-Form}. This gives us $c_D=-0.03(7)$ and $c_F=0.01(5)$, consistent with zero. In both scenarios, the axial coupling constants are consistent with the $n=1$ extrapolation. The discrepancies in the extrapolated results for all fitting forms we used is always at the level of one standard deviation. Therefore, we will assign an upper limit for the systematic error the same amount as the statistical error due to the extrapolation and keep the $n=1$ results as our central values.

The systematic errors due to finite-volume effects are expected to be small. LHPC calculated the nucleon axial coupling constant using a chiral extrapolation with finite-volume correction~\cite{Edwards:2005ym}; less than a 1\% effect is observed. Furthermore, finite-volume effects (including the $\Sigma$ and $\Xi$ baryons) are also estimated in Ref.~\cite{Detmold:2005pt}, using heavy baryon chiral perturbation theory. The finite-volume effects come in at a magnitude no larger than $10^{-4}$. Thus, we take 1\% to be an upper bound for the finite-volume systematic error in our calculation. Such effect is negligible given our statistical errors  and systematics due to chiral extrapolation.

The final source of systematic errors is the continuum extrapolation. Precise estimate of such errors requires computations at several lattice spacings. However, we can estimate the discretization errors which are $O(a^2\Lambda_{\rm QCD}^2)$. Taking  $\Lambda_{\rm QCD} \approx 300$~MeV and the value of our lattice spacing, $a^{-1} \sim~1588~{\rm MeV}$~\cite{Bernard:2001av}, one expects to have such a systematic error of the order of 4\%.

\section{Conclusion}\label{sec:conclusion}

In this work, we calculate the axial coupling constants for $\Sigma$ and $\Xi$ strange baryons using lattice QCD for the first time. We do the calculation using 2+1-flavor dynamical configurations with pion mass as light as 350~MeV. We discussed various potential chiral extrapolations and various sources of systematic errors. We conclude that $g_{A}=1.18(4)_{\rm stat}(6)_{\rm syst}$, $g_{\Sigma\Sigma}=0.450(21)_{\rm stat}(27)_{\rm syst}$ and $g_{\Xi\Xi} = -0.277(15)_{\rm stat}(19)_{\rm syst}$. In addition, the $SU(3)$ axial coupling constants are estimated to be $D=0.715(6)_{\rm stat}(29)_{\rm syst}$ and $F=0.453(5)_{\rm stat}(19)_{\rm syst}$. The axial charge coupling of $\Sigma$ and $\Xi$ baryons are predicted with significantly smaller errors than estimated in the past.

{\bf Acknowledgements:} The authors thank Martin Savage for motivating the project, W.~Detmold and C.~-J.~D.~Lin for the Mathematica notebook with their results of Ref.~\cite{Detmold:2005pt} and helpful discussion on further details. We thank the LHPC and NPLQCD collaborations for some of  the light and strange quark propagators. These calculations were performed using the Chroma software suite~\cite{Edwards:2004sx} on clusters at Jefferson Laboratory using time awarded under the SciDAC Initiative. This work is supported by Jefferson Science Associates, LLC under U.S. DOE Contract No. DE-AC05-06OR23177. The U.S. Government retains a non-exclusive, paid-up, irrevocable, world-wide license to publish or reproduce this manuscript for U.S. Government purposes. KO is supported in part by  the Jeffress Memorial Trust grant J-813, DOE OJI grant DE-FG02-07ER41527 and DOE grant DE-FG02-04ER41302.


\end{document}